\documentclass{aa}
\usepackage{txfonts}
\usepackage{graphicx}
\begin{document}
\def\nodata{~--}
\def\be{Be\,29}
\def\lesssim{\mathrel{\hbox{\rlap{\hbox{\lower4pt\hbox{$\sim$}}}\hbox{$<$}}}}
\def\gtrsim{\mathrel{\hbox{\rlap{\hbox{\lower4pt\hbox{$\sim$}}}\hbox{$>$}}}}
\newcommand{\ebv}{$0.093$}
\newcommand{\fehraw}{$-1.04 \pm 0.18$}
\newcommand{\fehcor}{$-0.74 \pm 0.18$}
\newcommand{\distmod}{$15.9 \pm 0.1$}

\newcommand{\evh}{\bf}
\newcommand{\cita}[1]{{#1}}
   \title{Radial velocities and membership of stars in the old, distant  
   open cluster Berkeley 29\thanks{
   Based on observations made with the Italian Telescopio Nazionale Galileo
   (TNG) operated on the island of La Palma by the Centro Galileo Galilei of 
   the INAF (Istituto Nazionale di Astrofisica) at the Spanish Observatorio 
   del Roque de los Muchachos of the Instituto de Astrofisica de Canarias
   }}
   
   \author{A. Bragaglia,\inst{1}
          E.V. Held\inst{2}
	  \and
	  M. Tosi\inst{1}}

   \offprints{A. Bragaglia}

   \institute{INAF -- Osservatorio Astronomico di Bologna,
     via Ranzani 1, 40127 Bologna (Italy),
              \email{angela.bragaglia@bo.astro.it, monica.tosi@bo.astro.it}
          \and
           INAF -- Osservatorio Astronomico di Padova, 
	     vicolo Osservatorio 5, 35122 Padova (Italy),
             \email{held@pd.astro.it}
             }

   \date{}

\abstract{
Multi slit spectroscopy at the Telescopio Nazionale Galileo
was employed to measure radial velocities for 20 stars in the
direction of the old open cluster Berkeley 29,  the farthest
known in our Galaxy.  Membership information was derived for stars
along all the red giant branch, in particular near its tip, and on the
red clump. The sample of bona-fide cluster members was used to revise 
the cluster distance to $\sim 15$ kpc, on the basis of an
empirical comparison with the red clump in open clusters with known
distances.  A metallicity [Fe/H] = \fehcor\ was also estimated using the
colours of spectroscopically confirmed red giant stars.
   
   }

\authorrunning{Bragaglia et al.}
\titlerunning{Membership of stars in the open cluster Be\,29}
   \maketitle

\section{Introduction}
\label{sec:introd}

Galactic open clusters are particularly well suited to study the disk
of the Milky Way and its history of chemical enrichment, since their
distances, ages, and metal abundances can be determined with relative
ease.  Until a few years ago, however, the data on cluster ages and
metallicities were sparse and inhomogeneous, and only recently
significant efforts have been dedicated to build large and well
studied cluster samples (e.g. Janes \& Phelps 1994; Friel 1995; Twarog
et al. 1997; Carraro et al. 1998).  Our group is undertaking
a long-term photometric project to derive a homogeneous set of ages,
metallicities, reddenings, and distances for a sample of open clusters
at various Galactocentric positions. To this end, we use the
colour-magnitude diagram (CMD) synthesis technique (see Tosi et
al. 1991).  Results have been published so far for 15 clusters
(e.g., Andreuzzi et al. 2004; Kalirai \& Tosi 2004;
Tosi et al. 2004 and
references therein).  This effort is complemented by accurate abundance
determinations derived by high-resolution spectroscopy (Bragaglia et al.
2001a,b; Carretta et al. 2004).

Age and distance determinations using the synthetic CMD method are
essentially based on reproducing the properties of the red clump,
of the main sequence (MS) and of its turn-off 
(MSTO).  For open clusters, which are usually heavily contaminated
by foreground and background stars, the separation of cluster members
and field interlopers becomes a prerequisite to properly derive the
cluster parameters.  In particular, it is important to know which
stars are real cluster members belonging to the red giant branch
(RGB), the red clump, and the MSTO in the
cluster CMD.  Membership assessment is also important to select
targets for follow-up high-resolution spectroscopy to measure detailed
chemical abundances.

Multi object spectroscopy at medium resolution represents an efficient
way to select samples of confirmed cluster stars in the direction of
Galactic open clusters on the basis of their radial velocities.  A
program has therefore been started to establish the membership of
stars in selected open clusters, in particular of stars in the crucial
evolutionary phases, by measuring their radial velocities using
multi object spectroscopy at the Telescopio Nazionale Galileo.

   \begin{figure}[ht]
   \centering
   \includegraphics[width=8.8cm]{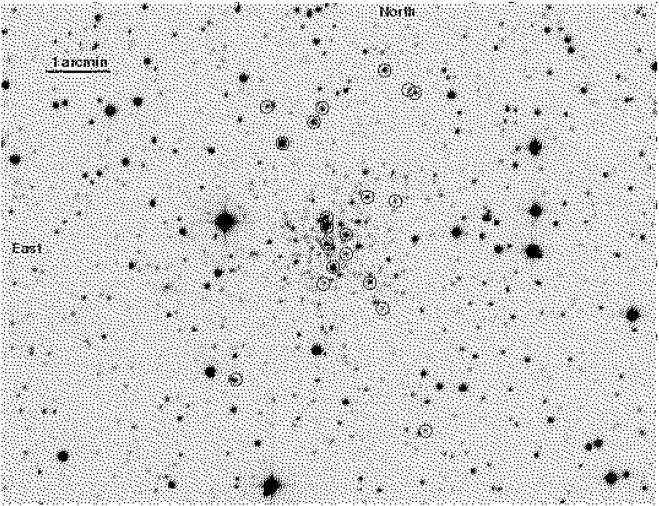}
   \caption{Map of the field around \be: the observed stars are marked
   by {\it circles}.  North is up, East to the left.}
   \label{Figmap}
   \end{figure}

We present here new results for the old, distant open cluster
Berkeley~29 (hereafter \be), located towards the Galactic Anticentre:
RA(2000) = $06^h 53^m 02^s$, DEC(2000) = $+16^o 56\arcmin 20\arcsec$,
$l = 197^o.98$, $b = 8^o.03$.
The photometric catalog is based on the CMD of Kaluzny (1994) and
 our $BVI$ data acquired with the CCD camera at the 1.5m
ESO/Danish telescope and the SuSI2 imager at the NTT ESO telescope at
La Silla, Chile (Tosi et al. 2004).
In his paper,  Kaluzny (1994) derived an age of about 4 Gyr for \be, and a
distance from the Sun of 10.5 kpc, making this the farthest known open cluster
(R$_{\rm GC} \gtrsim $ 18 kpc), hence a crucial object to study the radial
variations of disk properties.  Carraro \& Baume (2003) claimed that the old
cluster Saurer A is   even more distant from the Galactic centre,  but more
recent independent analyses (Carraro et al. 2004, Tosi et al. 2004) have
re-assessed that Be 29 appears indeed to be the outermost open cluster.

\be\ has a sparse RGB with 3 stars about 2 mag brighter than  the red clump
that could define the bright end of the RGB, if not its
tip. There is no published information on star membership, and even if this
cluster is not strongly contaminated by field stars, secure photometric
derivation of its distance, age, and metallicity requires an appropriate
cleaning of the CMD.

Spectroscopic confirmation of member stars is provided for the first
time in this paper. Our observations are presented in
Sect.~\ref{sec:obs}, and radial velocity measurements are discussed in
Sect.~\ref{sec:radvel}.  A brief discussion of the metallicity of \be\
based on the sample of spectroscopically confirmed stars is given in
Sect.~\ref{sec:discus}.  Our results are finally summarized in
Sect.~\ref{sec:sum}.

\section{Observations and data reduction}
\label{sec:obs}

\begin{table*}			    
\centering
\caption[]{
Data for the program stars.  Star numbers are taken  from our own
photometry
(Col.~1) and from Kaluzny (1994). The latter ID numbers are also used in the
Database for Stars in Open Clusters (BDA, Mermilliod 1995).  The
$B$, $V$, and $I$ magnitudes are from Tosi et al. (2004), and
equatorial coordinates are at J2000  (units of right ascension  are hours,
minutes, and seconds, and units of declination are degrees, arcminutes, and
arcseconds).  The classification and evolutionary phase of each candidate
member star is given in the last column.}
\begin{tabular}{rccccccl}
\hline\hline\noalign{\smallskip}
ID &ID$_{\rm K}$ &$B$ &$V$ &$I$ &RA(2000) &DEC(2000)         & phase   \\
\noalign{\smallskip}\hline\noalign{\smallskip}
 1024  &  241   &16.067 &14.458 &12.795 &6:53:07.132 &16:57:12.67 &RGB-tip \\
  869  &  902   &18.644 &17.739 &16.735 &6:53:03.048 &16:55:28.19 &RGB    \\
  994  &  988   &16.120 &14.585 &13.007 &6:53:03.886 &16:55:15.49 &RGB-tip \\
  136  & 1168   &19.352 &18.495 &17.507 &6:53:00.699 &16:54:36.48 &RGB    \\
  626  &\nodata &19.403 &18.530 &17.400 &6:52:57.923 &16:52:41.28 &RGB   \\
   72  &\nodata &18.216 &17.232 &16.090 &6:52:58.390 &16:57:58.75 &RGB  \\
  949  & 1076   &17.244 &16.221 &15.103 &6:53:01.482 &16:55:01.46 &RGB    \\
  159  &  412   &17.625 &16.627 &15.574 &6:53:01.601 &16:56:21.11 &clump  \\
  258  &  556   &17.583 &16.603 &15.534 &6:53:04.362 &16:56:02.86 &clump  \\
  933  &\nodata &17.435 &16.447 &15.322 &6:53:04.486 &16:57:44.69 &RGB  \\
  634  &\nodata &18.802 &17.908 &16.871 &6:52:58.788 &16:58:01.16 &RGB  \\
  104  &  441   &19.014 &17.945 &16.931 &6:52:59.759 &16:56:17.03 &RGB    \\
  257  &  818   &17.578 &16.608 &15.548 &6:53:04.320 &16:55:39.37 &clump  \\
  718  & 1075   &18.579 &17.658 &16.575 &6:53:04.514 &16:55:00.67 & RGB   \\
  784  &\nodata &18.532 &17.650 &16.590 &6:53:08.116 &16:57:47.17 &RGB  \\
 1009  &  673   &16.004 &14.310 &12.570 &6:53:04.385 &16:55:53.93 &RGB-tip \\
  993  &\nodata &16.320 &15.444 &14.453 &6:53:00.380 &16:58:20.27 & field \\
 1417  &\nodata &99.990 &14.818 &13.978 &6:53:05.051 &16:57:32.32 & field \\
  456  &\nodata &17.096 &16.350 &15.410 &6:53:10.436 &16:53:30.44 & field \\
  973  &  761   &16.322 &15.659 &14.899 &6:53:03.052 &16:55:46.16 & field \\
\noalign{\smallskip}\hline
\label{tabDATA}			     
\end{tabular}

\end{table*}

Intermediate resolution multi object spectroscopy of stars in \be\ was
obtained using the MOS mode of the low-resolution spectrograph
D.O.Lo.Res. at the Telescopio Nazionale Galileo (TNG) on Roque de los
Muchachos, Canary Islands, Spain.  The VPH grism centered on the
H$\alpha$ spectral region was used, yielding a resolution of 1.38 \AA\ 
(R=4750) with 1\farcs1 (200 $\mu$m) slitlets, and a spectral range of
about 680 \AA\ around H$\alpha$.
\be\ was observed in service mode on the nights of Jan. 11, 21, and
22, 2003. Spectra for 20 stars were obtained in two masks. Three 
exposures were obtained for each mask, with exposure
times ranging from 1800 to 2400 s.  A Ne lamp spectrum was acquired
immediately before and/or after each science frame for wavelength
calibration.

\begin{figure}
   \includegraphics[bb=50 175 530 640, clip=true, width=8.8cm]{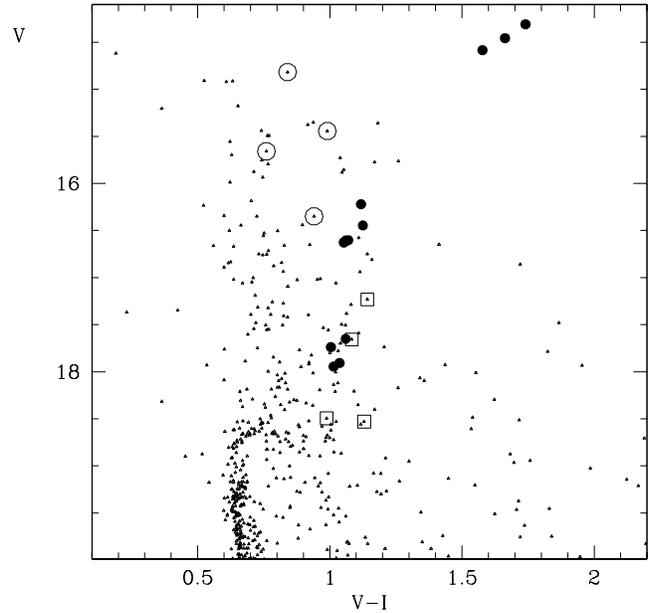}
   \caption{Colour-magnitude diagram for \be\ (Tosi et al. 2004), 
   with spectroscopically observed objects marked by larger
   symbols.  {\it Open circles} indicate field stars, {\it filled
   circles} are {\it bona-fide} members of \be, {\it open squares} are
   three non members, and one uncertain attribution.}
   \label{FigCMD}
   \end{figure}

Figure~\ref{Figmap} shows the field of \be, with the spectroscopically
observed stars marked by open circles.  Our target stars were chosen
on the basis of the observed CMD, using Tosi et al.'s (2004)
photometric data; stars in common with Kaluzny (1994) are
indicated in Table~\ref{tabDATA}, where all identifications,
magnitudes, and coordinates for the spectroscopically observed stars
are presented.  The target stars were selected along the RGB, the red
clump, and near the termination of \be \ RGB (the latter will be
referred to in the following as ``RGB-tip": this does not 
imply that they really define the tip, and is used only to
distinguish them from lower RGB stars).  We also included in the
sample 4 stars that, on the basis of their position in the CMD, are
likely field stars.  The positions of the observed stars in our $V,
V-I$ CMD (Tosi et al. 2004) are shown in Fig.~\ref{FigCMD}.

Two of the 3 RGB-tip stars had been previously observed by our group
with FEROS at the ESO 1.5m telescope (at R=48000); their spectra had
S/N ratio too low for deriving abundances, but high enough to measure
radial velocities of 24.76 and 24.14 km s$^{-1}$ for stars 
1009 and 1024, respectively (corrected for the solar system barycentric
motion).  These two stars with high-precision measurements 
 (having velocity uncertainties not larger than 1 km s$^{-1}$, 
including systematic error)
are useful as internal radial velocity standard stars.

The six MOS masks were reduced using standard packages in
IRAF\footnote{IRAF is distributed by the NOAO, which are operated by
AURA, under contract with NSF}.  The wavelength calibration residuals
have r.m.s. errors of about 0.01 \AA ~(less than 1 km s$^{-1}$).

\section{Radial velocities}
\label{sec:radvel}

\subsection{Velocity measurements}

\begin{table}	
\centering		    
\caption[]{ Results of radial velocity measurements.  The values for
the relative radial velocities ($\Delta v_r$, in column 2) are the average 
of the results of the cross correlation with star 1009 as template. 
Absolute radial velocities $v_r$ have been
obtained by adding the observed velocity of the template spectrum
(star 1009), measured from the H$\alpha$ line.  The quoted error
$\sigma_{v_r}$ represents the standard deviation of the mean.  
Radial velocities derived by cross correlation with the synthetic
spectrum as template are given in Col.~6, also corrected for the
barycentric motion. In the last column, a ``C'' flag indicates cluster
members, ``C:'' possible member, and ``F'' field stars.}
\begin{tabular}{rrrrcrl}
\hline\hline\noalign{\smallskip}
  ID & $\Delta v_r$  &$v_r$(cc,1) & $\sigma_{v_r}$ & $N_{\rm sp}$ & 
$v_r$(cc,2)     & \multicolumn{1}{c}{Flag}\\
\noalign{\smallskip}\hline\noalign{\smallskip}
1024 &  0.16   &  28.18 &  3.36  & 6 & 30.6   & \hspace{0.5em} C \\ 
 869 & $-$7.24 &  20.78 &  3.60  & 6 & 26.4   & \hspace{0.5em} C \\ 
 994 & $-$0.72 &  27.3  &  4.31  & 6 & 29.8   & \hspace{0.5em} C \\ 
 136 & 49.39   &  77.41 & 28.83  & 6 & 90.7   & \hspace{0.5em} F \\ 
 626 & 30.53   &  58.55 &  9.59  & 3 & 56.4   & \hspace{0.5em} F \\ 
  72 & 24.78   &  52.8  &  4.37  & 3 & 63.2   & \hspace{0.5em} F \\ 
 949 & $-$4.54 &  23.48 &  3.90  & 3 & 28.5   & \hspace{0.5em} C \\ 
 159 & $-$3.03 &  24.99 &  4.75  & 3 & 29.2   & \hspace{0.5em} C \\ 
 258 & $-$4.72 &  23.3  &  4.20  & 3 & 28.3   & \hspace{0.5em} C \\ 
 933 & $-$3.45 &  24.57 &  4.27  & 3 & 31.5   & \hspace{0.5em} C \\ 
 634 & $-$0.93 &  27.09 &  9.88  & 3 & 33.9   & \hspace{0.5em} C \\ 
 104 & $-$2.16 &  25.86 &  2.82  & 3 & 31.4   & \hspace{0.5em} C \\ 
 257 &  6.16   &  34.18 &  1.68  & 3 & 34.6   & \hspace{0.5em} C \\ 
 718 &$-$18.66 &   9.36 & 11.60  & 3 & 21.3   & \hspace{0.5em} C:\\ 
 784 &  4.20   &  32.22 &  6.28  & 3 & 36.1   & \hspace{0.5em} C \\ 
1009 &  0.0    &  28.02 &  0.52  & 6 & 28.2   & \hspace{0.5em} C \\ 
 993 & 61.23   &  89.25 &  8.04  & 6 & 93.9   & \hspace{0.5em} F \\ 
1417 & 60.67   &  88.69 &  3.92  & 6 & 93.3   & \hspace{0.5em} F \\ 
 456 &$-$36.29 &$-$8.27 &  2.90  & 6 & $-$7.9 & \hspace{0.5em} F \\ 
 973 &$-$27.72 &$-$0.30 &  4.99  & 3 &  5.4   & \hspace{0.5em} F \\ 
\noalign{\smallskip}\hline
\label{tabRV}			     
\end{tabular}

\end{table}

    \begin{figure*} \centering
    \includegraphics[bb=50 190 550 420, width=17.6cm]{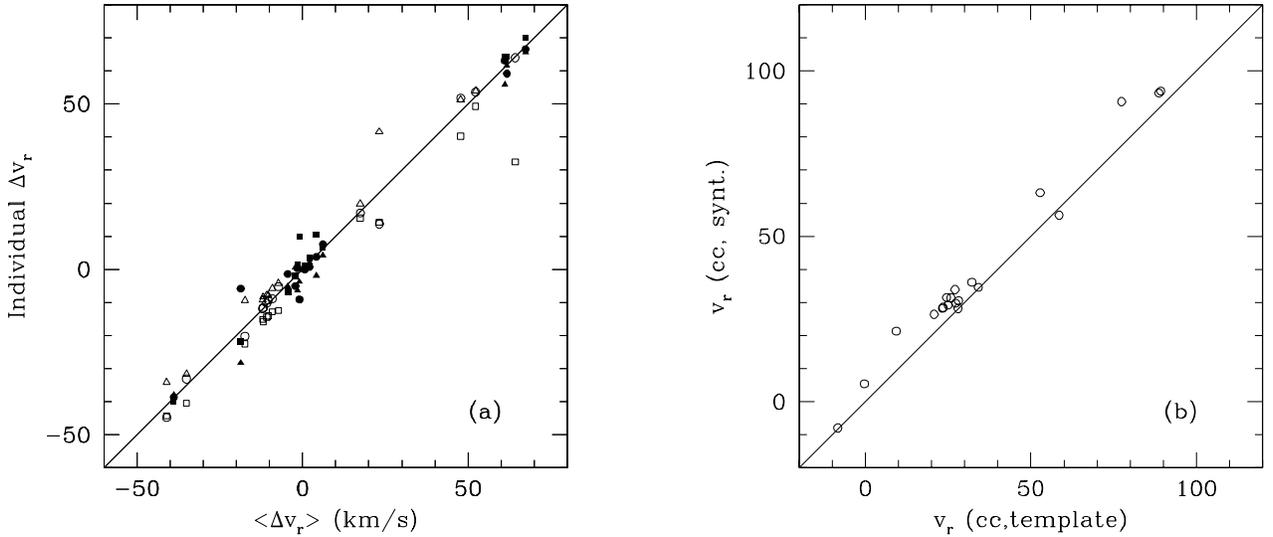}
   \caption{( a) A comparison of the average $\Delta v_r$ from
    cross-correlation with star 1009 with individual velocity measurements 
    for each
    stars (3 to 6 values available).  (b) Comparison of radial
    velocities obtained from
    cross-correlation with star 1009 and the synthetic template.} 
    \label{FigConf}
    \end{figure*}

 The radial velocities of stars in the field of \be\ were
measured in two ways. First, relative radial velocities ($\Delta
v_r$) were measured using the cross-correlation package FXCOR in
IRAF. The best spectrum of the brightest star in our sample (1009)
was used as a template; we also repeated the analysis using another
star (1024) and the results were very similar.  The
cross-correlation was computed in the spectral region 6520--6620 \AA,
chosen to avoid the presence of strong atmospheric absorption
features.  All the individual spectra were cross-correlated with the
template, obtaining 3 to 6 measures for each star. The formal errors
on the single $\Delta v_r$ provided by the cross-correlation method
range from $\sim 1$ km s$^{-1}$ to about 20 km s$^{-1}$ in the worst
case, with a typical value 4 km s$^{-1}$.
The mean value of the velocity relative to the template, $\Delta
v_r$, is given for each star in Table~2.  Since each mask set has a
different barycentric correction, all masks were referred to the
velocity system of a reference mask before averaging the radial
velocities.  The uncertainties were estimated from the standard
deviation of the measurements on the individual spectra,
$\sigma_{v_r}$.
The scatter of the single measurements is consistent with the errors
of the cross-correlation method.  A comparison of the individual
$\Delta v_r$ values for each star with the mean of the individual
measurements is presented in Fig.~\ref{FigConf}$a$.
Observed radial velocities $v_r$(cc) were obtained by adding the
observed velocity of the template spectrum (measured from H$\alpha$ as
below) to the relative velocities, and correcting for the barycentric
motion.  The results, given in Col.~3 of Table~2, provide the
reference data set of this paper.

For comparison, 
absolute radial velocities were also obtained from
direct measurement of the wavelength of absorption lines.  
In practice, radial velocities were measured from Gaussian fitting to
H$\alpha$ (the only line easily identified in all spectra) for each star,
using only the best mask.
The radial velocities measured from H$\alpha$ are also
given in Table~2, after correction for the barycentric motion, and the value
28.02 km s$^{-1}$ for star 1009 was used as zero point for the
cross-correlation derived values.

The zero point of our velocities agrees with previous high-resolution
FEROS measurements for stars 1009 and 1024 within 3-4 km
s$^{-1}$. We regard this as an excellent agreement given the limited
spectral resolution of our observations and the different sources of
systematic error with multislit spectroscopy, in particular centering
errors of the targets on the mask slits.

Finally, to further check our results the radial velocities were measured 
from cross-correlation with a synthetic template in the H$\alpha$ region for 
each star in all masks. 
The correlation between the radial velocities derived from the two
cross-correlations is good (Fig.~\ref{FigConf}$b$).  There is a systematic 
shift in the radial velocity scales of $\sim 4$ km s$^{-1}$ (median
difference), with the values estimated from cross-correlation with the
template stars being
lower. This small residual uncertainty is consistent with the typical
cross-correlation errors for our spectral resolution.  Anyhow, it is
of no consequence for discriminating cluster and field stars, which is
the main goal of this paper.  

\subsection{Membership}

   \begin{figure} \centering
   \includegraphics[width=9.8cm]{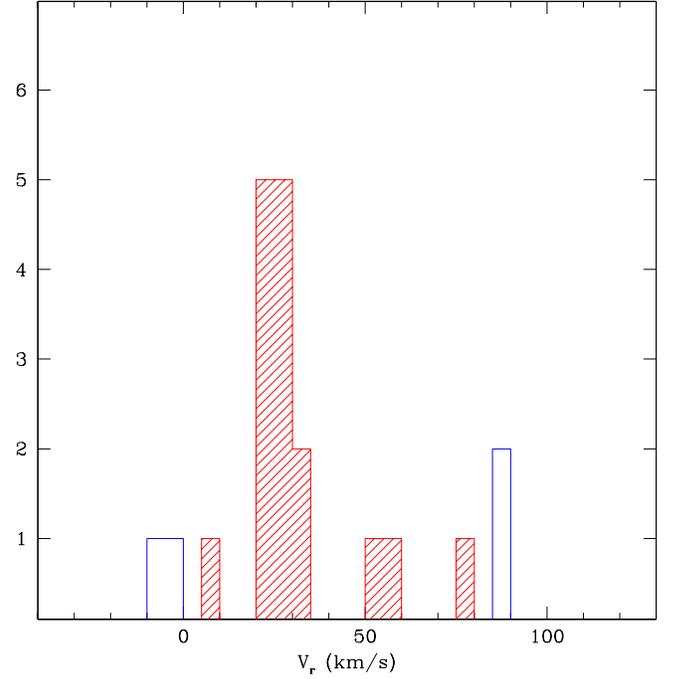}
   \caption{Histogram of the observed radial velocities  (Table
   \ref{tabRV}, Col. 3), using a bin of
    5 km s$^{-1}$. The hatched histogram is for supposed \be\ members, the
    open bins indicate the confirmed field stars.}
     \label{FigRV}
   \end{figure}

The histogram of the measured velocities (Fig.~\ref{FigRV}) shows a
very peaked distribution, with 4 stars in an uncertain position and
all the field stars far from the peak.  Our results for the membership
of stars in \be\ are summarized in the last column of Table~2.

Of the 16 candidate stars measured in \be, 12 turn out to be very
likely members, since their velocities are clustered around $\Delta
v_r = 0$. 
 One star, 718, has a radial velocity marginally consistent
with cluster membership, when considering cross-correlation with
star 1009, although the measurements have a large
scatter. However, both the velocity estimate from  H$\alpha$ on the best 
spectrum and cross-correlation with the synthetic template seem
to indicate that this is a probable cluster star.
Instead, objects  136 (the faintest in our sample), 626, and 72 appear to be
field stars.

\section{Discussion}
\label{sec:discus}

 Given its Galactic position, it is important to determine with precision 
the values of the metallicity and reddening, 
as well as distance, for \be.  
The age and metallicity of \be\ estimated by Kaluzny (\cita{1994})
from several photometric parameters, indicate that this cluster is
coeval to M\,67 ($\sim4$ Gyr old) and quite metal-poor, with a metal
abundance [Fe/H]$\approx -1.3$, and in any case much lower than that
of M\,67.
This conclusion largely rests upon the adopted reddening $E(B-V) =
0.21$ derived from the Burstein \& Heiles (\cita{1982}) reddening
maps.  By adopting an absolute magnitude $M_V=0.85$ for the red clump
and the above reddening, Kaluzny (\cita{1994}) obtained a distance
modulus $(m-M)_0 = 15.1$ for \be, equivalent to a heliocentric
distance of 10.5 kpc, making \be\ the most distant open cluster known.
On the other hand, Noriega-Mendoza \& Ruelas-Mayorga (\cita{1997}), by
applying a technique for the simultaneous determination of metallicity
and reddening, argued for a negligible reddening ($E(B-V) = 0.01$)
towards \be, and a higher metallicity ([Fe/H]$\sim -0.3$). If the
reddening is low, also the distance to \be\ must be revised.  In fact,
the infrared maps of Schlegel et al. (\cita{1998}) give a reddening
$E(B-V)$=\ebv\ in the direction of \be, lower than that adopted by
Kaluzny (\cita{1994}).  

 Very recently Tosi et al. (2004) have determined, using different
sets of evolutionary tracks and the synthetic colour-magnitude diagram
technique,
an age of 3.4 - 3.7 Gyr, an absolute distance modulus of 15.6 - 15.8
(which translate in a Galactocentic distance of 21.4 - 22.6 kpc),
and a reddening of 0.13 - 0.10; the metallicity, taken from the best fitting
tracks, has (formal) value of [Fe/H]$\sim -0.5$ or $-0.7$. 
Carraro et al. (2004) have obtained high resolution spectra of two
stars, and derived and abundance of [Fe/H] = $-0.44 \pm 0.18$, a reddening
of 0.08, an age of 4.5 Gyr, and a Galactocentic distance of 21.6 kpc.

To complement and further check these results, we will try to fully exploit our
present data set: the availability of a sample of spectroscopically
confirmed members of \be\ allows us to re-examine the distance and
metallicity of the cluster from their photometric data.
In particular, our sample of spectroscopically confirmed stars
includes 3 stars on the red clump allowing a secure determination of
the observed red clump magnitude and colour.
In the following we will assume the Schlegel et al. (1998) reddening
as our reference value, with an uncertainty of $\pm 0.05$ mag, since this
is in between the two most recent and precise determinations.
The mean magnitude of the red clump from the 3 stars turns out to be
$I = 15.55$ with a colour $V-I = 1.06$, in very good agreement with
previous determinations (Kaluzny \cita{1994}).
%
We note that the observed colour of the red clump in \be, once
corrected using the Schlegel et al. (\cita{1998}) value for the
reddening, is $(V-I)_0 = 0.95 \pm 0.06$. 
This is slightly blue for
an open cluster (Sarajedini \cita{1999}), but not exceptional at all
(Percival \& Salaris \cita{2003}) and goes in the direction of low
metallicity.  The reddening value adopted by Kaluzny (\cita{1994})
would imply a red clump as blue as $(V-I)_0 = 0.80$.

To estimate the distance to \be, we need to assume an absolute
magnitude for the red clump. There is still a debate about the
dependence of the red clump luminosity from metallicity and age of a
stellar population (see, e.g., Girardi \& Salaris \cita{2001}; Udalski
\cita{2000}).  To limit the impact of model-dependent assumptions we 
resorted to an empirical comparison with data for two 
well-studied open clusters (NGC 2506 and Mel 66), similar
enough to \be\ to represent a good comparison, and to minimize any
corrections we choose to apply.

   \begin{figure}[ht] \centering
   \includegraphics[width=9.cm]{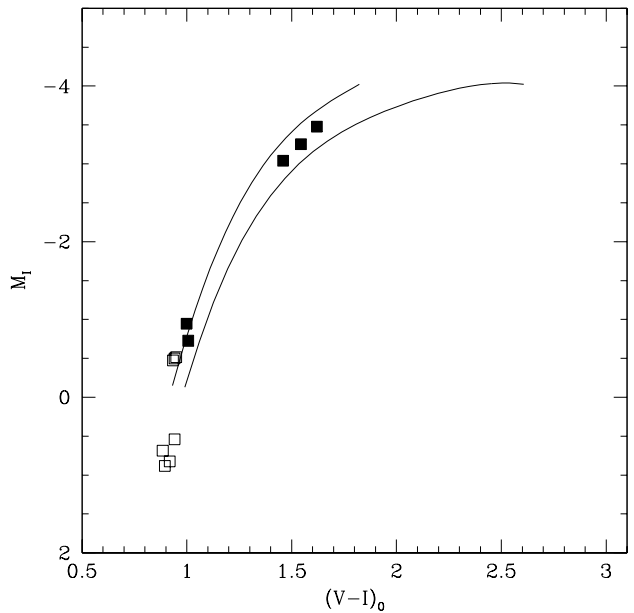}
   \caption{A comparison of spectroscopically identified members of
   \be\ with the fiducial lines of RGB stars in the Galactic globular
   clusters 47~Tuc and NGC\,1851 ([Fe/H] = $-0.71$ and $-1.29$
   respectively, from Da
   Costa \& Armandroff \cita{1990}).  For \be\ we have adopted a
   reddening $E(B-V) =$\ebv\ from Schlegel et al. (1998) and a
   corrected distance modulus $(m-M)_0 = 15.9$.  The stars used to
   estimate the metallicity of \be\ are shown as {\it filled
   squares}.}
   \label{FigMet}
   \end{figure}

NGC\,2506 has a metallicity [Fe/H]$=-0.39$ (Twarog et
al. \cita{1997}). Using a fit of theoretical main sequences and a
comparison to its almost-twin, but slightly metal-richer, NGC 2420,
Twarog et al. (\cita{1999}) estimated an age of about 2 Gyr, and from
the distance modulus they derived an absolute magnitude $M_I = -0.48$
and an extinction-corrected colour $(V-I)_0 = 0.94$ for the red clump.
Our first approach was to estimate the distance to \be\ under the
hypothesis that it is identical to NGC\,2506 (i.e., no age or
metallicity corrections were applied to its red clump magnitude). In
this case the uncorrected distance modulus is $(m-M)_I = 16.03$ and
the extinction-corrected distance modulus $(m-M)_0 = 15.86 \pm 0.09$.
Alternatively, we applied a correction to the luminosity of the red
clump in \be\ assuming a metallicity [Fe/H]$=-1.0$ and taking into
account the effects of a different age and metallicity.  The
correction, based on the models of Girardi \& Salaris (\cita{2001}),
turns out to be modest as it implies a luminosity $M_I = -0.50$ for
the red clump of \be.  The population correction due to the difference
in metallicity is nearly equal, within a few hundredths of magnitude,
to that derived using Popowski's (\cita{2000}) empirical calibration
of the red clump luminosity (his Eq.~4).  Using population corrections,
the corrected distance modulus becomes $(m-M)_0 = 15.90 \pm 0.09$.

These results are relatively independent of this particular choice
of the comparison cluster.  Using the open cluster Mel~66 as a
template, with the parameters derived by Sarajedini (1999), the
results are very similar.  Mel~66 is nearly coeval to \be, and has a
lower than solar metallicity ([Fe/H] $=-0.34$, Twarog et
al. \cita{1997}). The corrected distance modulus of \be\ turns out to
be $(m-M)_0 = 15.72 \pm 0.09$ assuming that the two clusters are
identical, and $(m-M)_0 = 15.85 \pm 0.09$ if we assume a metallicity
[Fe/H]$=-1.0$ for \be\ and apply population corrections.
Therefore our analysis implies that \be\ is 
located at $\sim 15$ kpc from the Sun.


Using this new distance estimate, and a reddening \ebv\ with a 0.05
mag uncertainty, we also measured the metallicity of \be\ by direct
comparison of $(V-I)$\ colours of its spectroscopically confirmed red
giant stars with the ridge lines of red giant branches in Galactic
globular clusters (Da Costa \& Armandroff \cita{1990}).
As shown in Fig.~\ref{FigMet}, the metallicity of \be\ appears to be
intermediate between those of NGC\,1851 and 47\,Tuc.
A quadratic interpolation of the mean colour difference between RGB
stars in \be\ and the ridge lines of globular cluster templates 
(see, e.g., Held et al. 1999 for details) yielded a new 
metallicity estimate [Fe/H]=\fehraw\ on the Zinn \& West
(1984) scale, not far from that found by Kaluzny (\cita{1994}).

This metallicity estimate assumes that \be\ is coeval to Galactic
globular clusters, which is certainly not the case.  Since \be\ has an
age lower than 4 Gyr (Kaluzny 1994, Tosi et al. 2004), its mean RGB colour 
is bluer than that of older Galactic cluster of the same metallicity.
We have estimated the effects of an age difference of $\sim 10$ Gyr
between \be\ and Galactic clusters by comparing the RGB sample with
isochrones of 4 and 14 Gyr and different metallicities (Girardi et
al. \cita{2000}). The shift in colour of the RGB of \be\ mimics a
metallicity difference of $\sim 0.3$ dex.  By applying this (model
dependent) correction, the colours of the spectroscopically confirmed
red giants imply an age-corrected metallicity [Fe/H]=\fehcor.

 In conclusion, also our analysis appears to exclude the low reddening and
high metallicity derived by Noriega-Mendoza \& Ruelas-Mayorga
(\cita{1997}), and favours a lower metallicity.
The value we find is lower than those found by Carraro et al. (2004) and
by Tosi et al. (2004), although consistent with the latter. Further 
high resolution spectroscopic data are clearly
needed to firmly establish the chemical abundances in Be29.

\section{Summary}
\label{sec:sum}

We have reported radial velocity observations of 20 stars in the
direction of the distant, old open cluster Be~29, obtained from
multi object spectroscopy at the TNG. On the basis of the radial
velocities, we found 12 confirmed cluster members,  1 probable
member, and three non members.
Thus we have been able to confirm that the 3 stars
on the brighter part of the RGB tip, 3 stars in the position of the
red clump, and 6 stars on the fainter RGB are indeed cluster members.

This information has been used to re-examine cluster properties such
as distance and metallicity using only spectroscopically confirmed
stars.  Our analysis used an empirical comparison of the magnitude and
colour of the red clump in \be\ and similar open clusters with
published data to derive a corrected distance modulus
$(m-M)_0$=\distmod, corresponding to $\sim 15$ kpc.

An estimate of metallicity based on the colours of red giants was
obtained by direct comparison with the RGB fiducial lines in Galactic
globular clusters, yielding an age-corrected metallicity
[Fe/H]=\fehcor, a value supporting the suggestion that \be\ is
moderately metal-poor.

Further constraints must be derived from stars near the MSTO, and this
will be the goal of new observations. With our study we have also
indicated candidates for detailed chemical analysis  based on further
high resolution spectroscopy of a larger sample of stars; 
this requires telescopes of the 10m class
for all stars except the three near the RGB tip.

\begin{acknowledgements}

 We thank L. Di Fabrizio and all the TNG telescope staff who ensured
completion of the observations, and E. Carretta who computed the
synthetic spectrum.  The cross-identification of stars was
done using software written by P. Montegriffo.  
We thank the referee, J. Kaluzny for the useful suggestions that helped
to make the paper clearer.
We acknowledge the use of the Database for Stars in Open Clusters
(http://obswww.unige.ch/webda/) developed by J.-C. Mermilliod.  This
research has made use of the NASA/IPAC Extragalactic Database (NED)
which is operated by the Jet Propulsion Laboratory, California
Institute of Technology, under contract with the National Aeronautics
and Space Administration.
\end{acknowledgements}


\end{document}